\documentclass{moriond}
\usepackage{mathrsfs,amsfonts,amsmath}

\newcommand{\Photo}{}

\begin{document}
\vspace*{4cm}
\title{SOME RECENT DEVELOPMENTS IN 5TH FORCE SEARCHES}

\author{Ephraim Fischbach$^{a,b}$, Dennis E. Krause$^{c,a}$, Megan H. McDuffie$^{a}$, and Michael J. Mueterthies$^{a}$}

\address{$^{a}$Department of Physics and Astronomy, Purdue University West Lafayette, IN 47907 USA \\
$^{b}$Snare, Inc., West Lafayette, IN 47906, USA\\
$^{c}$Department of Physics, Wabash College, Crawfordsville, IN 47933, USA}

\maketitle\abstracts{
Recently we drew attention to the fact that  most recent 5th force searches and tests of  the weak equivalence principle (WEP) utilize only one or two pairs of test samples.  We argue that, despite the great precision of these experiments, the lack of diversity of samples may mean they are unable to detect new composition-dependent forces, which requires the observation of a {\em pattern} in the data.  Such a pattern was observed in  the experiment by  E\"{o}tv\"{o}s, Pek\'{a}r, and Fekete (EPF), the  last high precision test of the WEP which used a significant number of different samples.  We advocate for new experiments utilizing a sufficient number and range of samples to either confirm or refute the pattern found in the EPF experimental data.}

\section{Introduction}
Beginning with Newton and continuing through the early 20th century, experiments searching for a composition-dependence of the force of gravity, i.e., violations of the weak equivalence principle (WEP), utilized a rich variety of materials.\cite{Newton,Bessel,EPF,Potter}  However, this is not true for modern, more precise experiments\cite{Dicke,Braginskii,Wagner,Touboul PRL} as seen in Table~\ref{WEP table}.  The signature of a composition-dependent force would be a {\em pattern} in the acceleration data of a WEP experiment which can only be observed when a sufficiently large number of different samples is used, as was done in the earlier tests of the WEP.   Such a pattern, a coupling to baryon number was, in fact, observed in the data of the most sensitive experiment which used more than three samples,\cite{Fischbach PRL}  the torsion balance experiment by E\"{o}tv\"{o}s, Pek\'{a}r, and Fekete (EPF).\cite{EPF}  Yet, despite a significant subsequent experimental effort, no compelling additional evidence for a new ``fifth force'' coupling to baryon number has been observed.  However, as we pointed out recently,\cite{EPF paper} the more contemporary experiments have only utilized one or two pairs of different materials.  

Here we  briefly review the work by Fischbach, et al., \cite{Fischbach PRL,Fischbach AoP} which discovered the pattern in the EPF data, and what the data can tell us about the force which might have caused it.  We also discuss how subsequent experiments might have failed to observed this force.  As an example, we show that the EPF experiment, and experiments by the E\"{o}t-Wash group,\cite{Wagner} have different sensitivities to new forces, this despite the fact they both use torsion balances.  Together, these observations build a compelling case for conducting new WEP experiments with sensitivities comparable to the most recent experiments, but which compare accelerations for a significantly broader range of samples.\cite{EPF paper}

\begin{table}[t]
\caption[]{Tests of the WEP with the method, the number of different samples and pairs of samples  used, and their sensitivity $\Delta \kappa = \Delta a/g$.  (The number of pairs for pendulum experiments are not given since it is the same as the number of samples.)}
\label{WEP table}
\vspace{0.4cm}
\begin{center}
\renewcommand{\arraystretch}{1.1}
\begin{tabular}{| l l l l c l l |}
\hline
{\bf Experiment} & \hspace{.1cm}  & {\bf Method}  &  \hspace{.1cm}  & {\bf \# of Samples} & \hspace{.1cm}  & {\bf Sensitivity (\boldmath $\Delta \kappa$)} \\ \hline
Newton\,\cite{Newton} (1687) && Pendulum && 9 && $\sim 10^{-3}$ \\
Bessel\,\cite{Bessel} (1832) && Pendulum && 10 & &  $\sim 2 \times 10^{-5}$\\
E\"{o}tv\"{o}s\,\cite{EPF} (1922) && Torsion Balance && 11 (10 pairs) & & $\sim  10^{-9}$ \\
Potter\,\cite{Potter} (1923)&& Pendulum && 7& &$\sim 3 \times 10^{-6}$ \\
\hline
Dicke\,\cite{Dicke} (1964) && Torsion Balance && 2 (1 pair) & & $\sim 10^{-11}$ \\
Braginskii\,\cite{Braginskii} (1972) && Torsion Balance&& 2 (1 pair)  & & $\sim 10^{-12}$\\
E\"ot-Wash\,\cite{Wagner} (2012) &&Torsion Balance && 3 (2 pairs) & & $\sim 10^{-13}$ \\
MICROSCOPE\,\cite{Touboul PRL} (2017) &&Satellite && 2 (1 pair)  && $\sim 10^{-15}$
\\ \hline
\end{tabular}
\end{center}
\end{table}

\section{The E\"otv\"os Pattern and Paradox}

\begin{figure}[t]
\centering\includegraphics[width=0.5\linewidth]{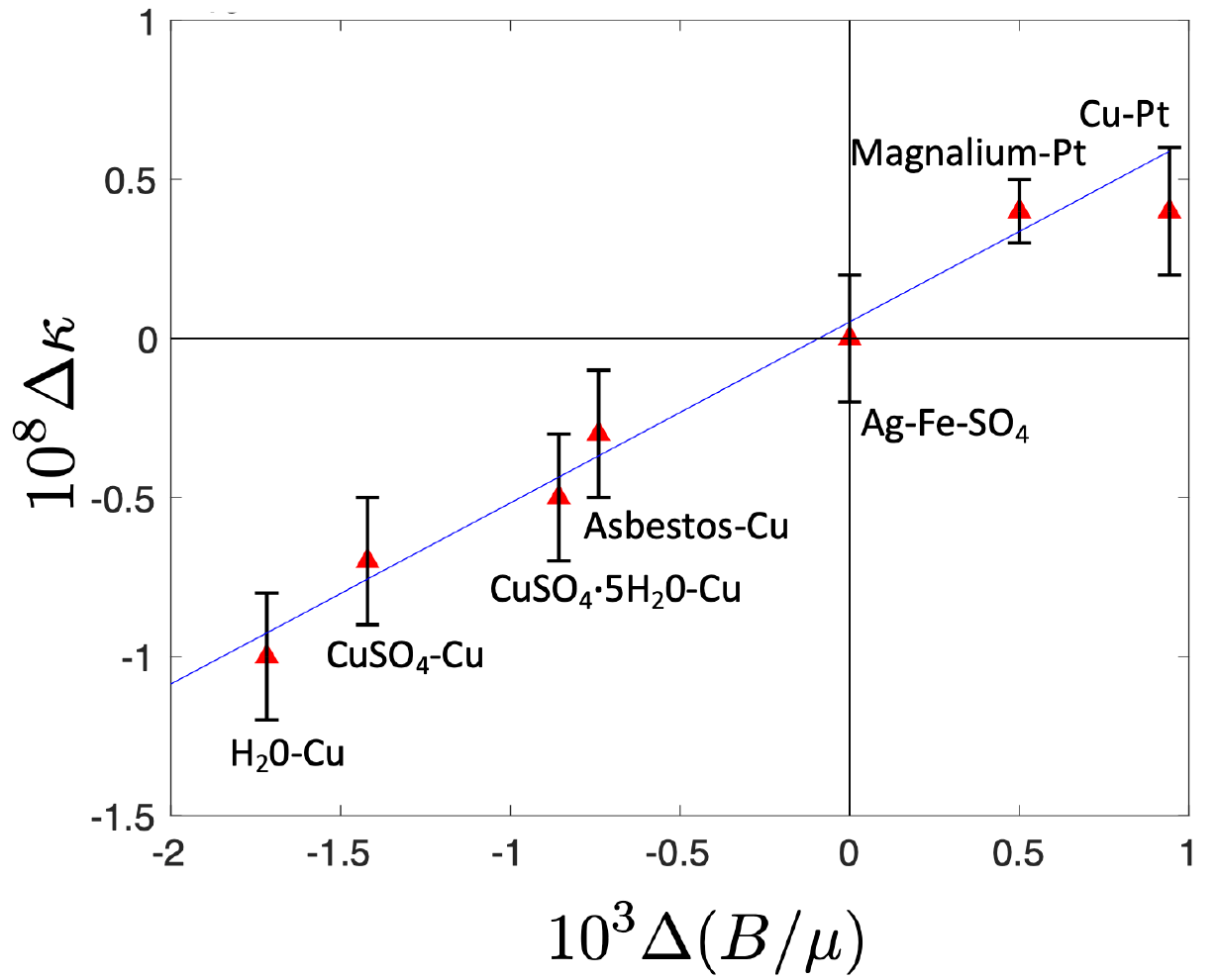}
 \caption{The E\"otv\"os Pattern: Data from the EPF experiment exhibit a  linear dependence of $\Delta\kappa = \Delta a/g$ on the baryon-to-mass ratio difference $\Delta(B/\mu)$ which passes through the origin.}
 \label{EPF figure}
\end{figure}

The last high-precision WEP experiment which used a significant number of different samples was conducted by  E\"{o}tv\"{o}s, Pek\'{a}r, and Fekete (EPF), the results of which  were published in 1922.\cite{EPF}  Motivated by a number of theoretical and experimental considerations, Fischbach and colleagues reanalyzed the EPF data, discovering evidence of a new Yukawa ``fifth force'' coupling to baryon number $B$ with range $\lambda$, \cite{Fischbach PRL}
\begin{equation}
V_{5}(r) \propto \frac{B_{1}B_{2}}{r}e^{-r/\lambda}.
\label{V5}
\end{equation}
Specifically, when the acceleration differences relative to gravity ($\Delta\kappa = \Delta a/g$) of the samples were plotted versus their baryon-to-mass ratio differences $\Delta(B/\mu$) ($\mu = m/m_{H}$, the sample mass relative to atomic hydrogen), one observes that the data fall along a line with a vanishing intercept  (Fig.~\ref{EPF figure}).    That is, if we write
\begin{equation}
\Delta\kappa = a\, [\Delta(B/\mu)] + b,
\end{equation}
a weighted least-squares fit to the EPF data finds that the slope $a$ and intercept $b$ are given by\cite{Fischbach PRL,EPF paper}
\begin{equation}
a  = (5.65 \pm 0.71) \times 10^{-6}, \phantom{space}
b  = (4.83 \pm 6.44) \times 10^{-10}.
\end{equation}
Artificially inflating the uncertainties to force the data to agree with the WEP only worsens the $\chi^{2}$ of the fit without affecting the slope.\cite{EPF paper} We are then led to the {\bf E\"otv\"os Paradox}: There is no evidence of any fault in the  experiment by EPF, or in the subsequent re-analysis of their data by Fischbach, et al., which revealed the pattern shown in Fig.~1.  Yet, no convincing evidence of a Yukawa force coupling to baryon number has been found in subsequent experiments.\cite{Wagner,Tino,Franklin,Fischbach Book,Adelberger,Will}  How is this possible?

\section{Towards a Resolution of the  E\"otv\"os Paradox}

We take the first steps towards resolving the E\"otv\"os paradox by noting that the pattern shown in Fig.~\ref{EPF figure} does not give any indication of the spatial dependence of the interaction which may be causing the pattern.  While recent experiments have set very stringent limits on new Yukawa interactions of the form given by Eq.~(\ref{V5}), this does not necessarily exclude other non-Yukawa 
forces  coupling to baryon number.  The E\"otv\"os pattern can be produced by any force  proportional to baryon number.   If we write the formula for the force on a test body as $\vec{F}_{B} = B \vec{\mathbb{F}}$, where $\vec{\mathbb{F}}$ is a baryon force field, the  difference in accelerations of two test bodies  relative to their gravitational accelerations is
\begin{equation}
\frac{\Delta\vec{a}}{g}=  \Delta \!\left(\frac{B}{\mu}\right)\frac{\vec{\mathbb{F}}}{m_{H}g} \equiv \Delta \! \left(\frac{B}{\mu}\right)\vec{\cal F},
\label{a/g}
\end{equation}
where $\vec{\cal F}$ is the dimensionless force field.  In the original fifth force hypothesis by Fischbach et al,  the source of the E\"otv\"os pattern was the Earth with the Yukawa interaction given by Eq.~(\ref{V5}).  Then the dimensionless fifth force field would be \cite{Fischbach PRL}
\begin{equation}
{\cal F} = {\cal F}_{5} =  \frac{f^2\epsilon(R_{\oplus}/\lambda)}{G m^2_H} \left(\frac{B_\oplus}{\mu_\oplus}\right),
\end{equation}
where $f$ is the fifth force charge, $B_{\oplus}/\mu_{\oplus} \simeq 1$ is the average baryon-to-mass ratio of the Earth, and $\epsilon(R_{\oplus}/\lambda)$ is a constant arising from integrating over the volume of the Earth.

While a large number of experiments have found no evidence for a Yukawa fifth force, this does not exclude the possibility that another model for a fifth force may exist that can produce the E\"otv\"os pattern, and yet would not have been seen in these more recent experiments.  There are several reasons for this.  First, as shown above, any force with a linear coupling in $B$ will produce the E\"otv\"os pattern, not just a Yukawa potential.  Second, none of the newer experiments utilize more than two pairs of different samples and so cannot convincingly reproduce the E\"otv\"os pattern.  By investing great effort in maximizing precision at the price of using only two or three samples, it is possible that an unexpected new force might be unintentionally zeroed-out by misinterpreting it as an irrelevant background effect.  Finally, one usually  optimizes an experiment to maximize its sensitivity to an {\em expected} signal, but possibly at the expense of reducing its sensitivity to an {\em unexpected} signal.

To investigate the latter effect, Mueterthies compared the sensitivity of the EPF experiment with the E\"ot-Wash (EW) group's experiment,\cite{Wagner}  both  of which use torsion balances, to a very general dimensionless fifth force.\cite{Mueterthies}  Specifically, he assumed both EPF and EW  torsion balances where acted upon by a force whose  $i$th component was given by $F_{i} = B\left(\mathbb{F}_{i} + \sum_{j}\mathbb{D}_{ij}r_{j}\right)$,
where $\mathbb{F}_{i}$ and $\mathbb{D}_{ij}$ are components of vector and tensor force fields, and $r_{j}$ is the $j$th component of the sample's position relative to the torsion balance's pivot point.
Mueterthies  then showed that the differential accelerations produced by this force on the EW and EPF apparatuses were
\begin{eqnarray}
\Delta\kappa_{EW} & \equiv & \left(\frac{\Delta a}{g}\right)_{\rm EW} = \sqrt{2} \Delta\!\left(\frac{B}{\mu}\right)\sqrt{\left({\cal F}_{x}+{\cal F}_{z}\sin\beta\right)^{2}+{\cal F}_{y}^2},  \label{kappa EW} 
\\
\Delta\kappa_{EPF} & \equiv & \left(\frac{\Delta a}{g}\right)_{\rm EPF}  =\Delta\!\left(\frac{B}{\mu}\right)
 \left[
 \left({\cal F}_{z}\sin\beta + {\cal F}_{x}+\frac{\nu}{\sigma}{\cal F}_{y}\right) \right. \nonumber \\
& & \mbox{} - L\frac{\nu}{2\delta}({\cal D}_{yy}-{\cal D}_{xx}) 
 + h\left({\cal D}_{xz}+\frac{\nu}{\sigma}{\cal D}_{yz}\right) 
\nonumber\\
& & \left. - h\sin\beta\left({\cal D}_{xx}+\frac{\nu}{\sigma}{\cal D}_{yx}-{\cal D}_{zz}\right)\right], \label{kappa EPF} 
\end{eqnarray}
respectively.  Here ${\cal F}_{i} = \mathbb{F}_{i}/m_{H}g$ and ${\cal D}_{ij} = \mathbb{D}_{ij}/m_{H}g$ are the components of the normalized force fields, and $\beta$ is the angle of the torsion fiber relative to the vertical.  The other parameters $\nu$, $\sigma$, $L$, $h$ characterize the dimensions of the EPF experimental setup.\cite{Mueterthies}

The differences between $\Delta\kappa_{\rm EW}$ and $ \Delta\kappa_{\rm EPF}$  given by Eqs.~(\ref{kappa EW}) and (\ref{kappa EPF}) are clearly evident. While both experiments are sensitive to the fifth force components ${\cal F}_{i}$, their functional dependence differs significantly.  Furthermore, the EPF experiment is sensitive to a fifth force that depends on ${\cal D}_{ij}$ (force gradients), while these  terms are absent from $\Delta\kappa_{\rm EW}$.  This vividly illustrates an important point:  while the EPF and EW experiments both use torsion balances, their different configurations of test masses and methodologies make them potentially sensitive to different signals.    It is thus clearly possible for a new force to be detected by one experiment and be missed by the other,  as demonstrated by a simple model given by Mueterthies.\cite{Mueterthies}

\section{Conclusions}

We have argued that, despite their great precision, modern WEP tests utilize too few pairs of samples to detect a pattern such as seen in the EPF experimental data.   New high-precision WEP experiments using pairs of samples with the widest possible range of $B/\mu$ are needed to search for, or exclude, the E\"otv\"os pattern.

\section*{Acknowledgments}

We thank J. T. Gruenwald and C. Y. Scarlett for useful discussions on this work.

\end{document}